\documentclass[fleqn,usenatbib]{mnras}
\usepackage[T1]{fontenc}

\DeclareRobustCommand{\VAN}[3]{#2}
\let\VANthebibliography\thebibliography
\def\thebibliography{\DeclareRobustCommand{\VAN}[3]{##3}\VANthebibliography}

\usepackage{graphicx}	
\usepackage{amsmath}	
\usepackage{amssymb}	

\def\hi{H\,{\sc i}}

\def\kms{km~s$^{-1}$}

\def\msun{$M_{\odot}$}

\def\mueff{$\left< \mu_\mathrm{eff}\right>$}
\def\mueffi{$\left<\mu_\mathrm{eff}^\mathrm{I}\right>$}
\def\mueffirac{$\left<\mu_\mathrm{eff}^\mathrm{3.6}\right>$}
\def\jm{$j_*$--$M_*$}
\def\simba{\textsc{Simba}}
\def\irac{3.6~$\mu$m}

\title[The \simba\ mass-spin-surface brightness relation]{Using the Simba cosmological simulations to measure the planar relation between stellar specific angular momentum, mass and effective surface brightness}

\author[E. Elson]{
E. Elson,$^{1}$\thanks{E-mail: eelson@uwc.ac.za (EE)}
\\
$^{1}$Department of Physics \& Astronomy, University of the Western Cape, Robert Sobukwe Rd, Bellville, 7535, South Africa\\
}

\date{Accepted 2024 September 11. Received 2024 September 11; in original form 2024 June 6}

\pubyear{2024}

\begin{document}
\label{firstpage}
\pagerange{\pageref{firstpage}--\pageref{lastpage}}
\maketitle

\begin{abstract}
Stellar mass and specific angular momentum are two properties of a galaxy that are directly related to its formation history, and hence morphology.  In this work, the tight planar relationship between stellar specific angular momentum ($j_*$), mass ($M_*$) and mean effective surface brightness (\mueff) that was recently constrained using ALFALFA galaxies is measured more accurately using galaxies from the \simba\ cosmological simulation. The distribution of 179 \simba\ galaxies in $\log_{10}j_*$~-~$\log_{10}M_*$~-~\mueff\ space is shown to be very tightly planar with $j_*\propto M_*^{0.694}$ and the distribution of perpendicular distances between the galaxies and the plane being approximately Gaussian with $\mathrm{rms}=0.057$~dex.  The parameterised distribution is used with existing $j_*$ and \mueff\ measurements of 3~607 ALFALFA galaxies and 84 SPARC galaxies to reliably predict their published stellar masses to within $\sim 0.1$ to 0.2~dex over several decades of stellar mass.  Thus, this work presents a new method of easily generating accurate galaxy stellar mass estimates for late-type galaxies and provides a new measurement of the fundamental link between galaxy morphology, mass and angular momentum.
\end{abstract}


\section{Introduction}
In a $\Lambda$CDM Universe, the hierarchical buildup of galaxies implies that a galaxy's angular momentum content is primarily acquired through the tidal torques exerted during mergers and interactions with other structures (e.g., \citealt{hoyle_1949, peebles_1969}). This process results in the transfer and redistribution of angular momentum, influencing the rotation and spin of galaxies as they evolve and grow.  For dark matter haloes in a standard $\Lambda$CDM Universe, $j_\mathrm{h}\propto \lambda M_\mathrm{h}^{2/3}$, where $j_\mathrm{h}$ is a halo's specific angular momentum and $M_\mathrm{h}$ its mass.  By considering the ratio of a galaxy's stellar mass to that of its halo, as well as the ratio of the average specific angular momentum in stars in the galaxy to that of its halo, its total stellar specific angular momentum can be shown to be $j_*\propto M_*^{2/3}$, where $M_*$ is its stellar mass\footnote{See \citealt{posti_2018a} for a concise derivation of full the equation.}. 

More than 40 years ago, \citet{fall_1983} provided empirical evidence for the existence of a tight power-law relationship between a galaxy's stellar specific angular momentum and its mass: $j_*=\beta M_*^{\alpha}$, where $\alpha\approx 0.6$ and $\beta$ is related to galaxy morphology.  Since then, many other studies have confirmed the power-law relationship and have further constrained its parameters (e.g., \citealt{romanowsky_fall_2012, fall_romanowsky_2013, OG14, cortese_2016, Elson_2017, posti_2018b}, and others).  The value of $\alpha$ is usually close to 2/3 while galaxy morphology - or suitable proxies thereof - seems to affect the value of $\beta$.  

Several investigators have explicitly studied how the scatter seen in the \jm\ relation is related to various proxies of galaxy morphology.  Considering a galaxy's  gas fraction, $f_\mathrm{gas}$\footnote{$f_\mathrm{gas}=1.33M_\mathrm{HI}/M_*$},  \citet{mancera_pina_2021b} fitted 2D planes to the distributions of galaxies in $\log_{10}j_*$~--~$\log_{10}M_*$~--~$\log_{10}f_\mathrm{gas}$ space and reported that at fixed $M_*$ galaxies with higher $f_\mathrm{gas}$ typically also have higher $j_*$.  More recently, \citet{elson_2024a}  used a large sample of 3607 galaxies from the Arecibo Legacy Fast ALFA (ALFALFA) survey \citep{haynes_2018} to demonstrate for them the existence of a very tight planar distribution in $\log_{10}j_*$~--~$\log_{10}M_*$~--~\mueffi\ space, where \mueffi\ - the magnitude measure (in the SDSS~$I$-band) of the average flux surface density of a galaxy within its effective radius - was used as a proxy for galaxy morphology.  Projecting \mueffi-selected sub-samples of the 3607 galaxies onto $\log_{10}j_*$~--~$\log_{10}M_*$ space yielded the tightest \jm\ relations measured to date over the stellar mass range $8\lesssim \log_{10}(M_*/M_{\odot})\lesssim 11$.  \footnote{ Note that the \citet{elson_2024a} study was based on spatially-unresolved \hi\ imaging while the studies of \citet{posti_2018b} and \citet{mancera_pina_2021b} were based on spatially-resolved \hi\ maps.}

The goal of the present study is to use galaxies from the \simba\ cosmological galaxy formation simulations \citep{dave_2019} to better constrain the distribution of galaxies in $\log_{10}j_*$~--~$\log_{10}M_*$~--~\mueff\ space.  A parameterisation of the planar distribution of \simba\ galaxies in the 3D space is shown to serve as a useful tool for accurately estimating the stellar masses of real galaxies for which there are available measurements of their global properties that can be used to constrain their $j_*$ and \mueff\ quantities. 

The layout of this paper is as follows.  Section~\ref{simba_sims} briefly introduces the \simba\ simulations.  The data products generated and used in this study are discussed in Section~\ref{sec:measurements}.  The sample of \simba\ galaxies used in this study is discussed and presented in Section~\ref{gal_sample}.  The study's main results appear in Section~\ref{sec:results} while a demonstration of their utility to predict the stellar masses of real galaxies is given in Section~\ref{mstarpredicting}.  Finally, Section~\ref{sec:summary} offers a summary of this study.
\section{\textsc{Simba} simulations}\label{simba_sims}
The \textsc{Simba} cosmological galaxy formation simulations provide all of the data used in this study.  \textsc{Simba} is run with \textsc{Gizmo’s} meshless finite mass hydrodynamics.  The simulation includes prescriptions for various star formation processes including blackhole growth.  This work utilises the $z=0$ simulation box of the highest resolution, consisting of $512^3$ dark matter particles and $512^3$ gas elements with periodic volume of size 25~$h^{-1}$~Mpc$^3$.  The associated mass resolutions are $1.2\times 10^7$~\msun\ and $2.28\times 10^6$~\msun\ for the dark matter particles and gas elements, respectively.  Rather than explicitly model physical processes that control the cold phase of the interstellar medium, \textsc{Simba} uses various prescriptions to transform ionised gas into atomic and molecular phases.  Star formation is modelled using an H$_2$-based \citet{schmidt_1959} relation that uses H$_2$ density and dynamical time.  To model stellar evolution, a \citet{chabrier_2003} stellar initial mass function is assumed and a 6D friends-of-friends algorithm is used to group star particles into galaxies.  

A total of 5218 galaxies are associated with the 25~$h^{-1}$~Mpc$^3$ simulation volume. For \textsc{Simba}, these galaxies have had their properties computed using \textsc{Caesar}\footnote{https://caesar.readthedocs.io/en/latest/}, which is a particle-based extension to \textsc{YT}\footnote{https://yt-project.org} which itself is a Python package for visualising volumetric data.  Photometry for \textsc{Simba} galaxies is performed using \textsc{PyLoser}\footnote{https://pyloser.readthedocs.io/en/latest/}, which is a Python version of \textsc{Loser} package \citep{dave_2017} that computes line-of-sight dust column densities through the gas of a given galaxy.

\section{Measurements}\label{sec:measurements}
This study aims to measure the distribution of galaxies in $\log_{10}j_*$~-~$\log_{10}M_*$~-~\mueff\ space.  A galaxy's stellar specific angular momentum is approximated as 
\begin{equation}\label{j_approx}
\tilde{j_*}=2V_\mathrm{circ}R_\mathrm{d},
\end{equation}
where $\tilde{j_*}$ is the approximation of $j_*$,  $V_\mathrm{circ}$ is the circular velocity and $R_\mathrm{d}$ is the exponential disc scale length.  If a spiral galaxy's stellar mass distribution is well-described by a thin exponential disc, and if a radially constant rotation curve is assumed, $\tilde{j_*}$ provides an excellent approximation of $j_*$.  This fact is well-demonstrated by \citet{romanowsky_2012} for a sample of nearly 100 nearby galaxies.  From this point forward, $j_*=\tilde{j_*}$ is assumed.  

Assuming a galaxy's stellar mass to be exponentially distributed with radius, the disc scale length is taken to be $R_\mathrm{d}=R_\mathrm{eff}/1.68$, where $R_\mathrm{eff}$ is its effective radius from the \textsc{Caesar} catalogue.  Circular velocity is calculated as ${V_\mathrm{c}=W_{50}/ 2\sin i}$, where $W_{50}$ is a measure of the velocity width of its \hi\ line profile, in units of \kms, and $i$ is the \hi\ disc inclination angle.  This conversion of velocity width to circular velocity assumes corrections for pressure-supported motions to be negligible.  \citet{mancera_pina_2021a} showed that for the mass regime of the galaxy sample used in this work, circular velocity is very close to the rotation traced by the gas.

In this study, each galaxy's $W_{50}$ is measured from an \hi\ line profile that is generated by spatially integrating the flux in the channels of an \hi\ data cube generated for the galaxy. The MARTINI (Mock APERTIF-like Radio Telescope Interferometry of the Neutral ISM) package \citep{martini_1, martini_2, martini_3} is used in this work to create synthetic resolved \hi\ line data cubes of the \textsc{Simba} galaxies.  All cubes are given an RA-DEC pixel size of 4~arcsec and a channel width of 5~\kms.    A distance of 4~Mpc is used for all galaxies while a Gaussian point-spread function of half-power width 12~arcsec is used to spatially smooth the data.  To produce cubes essentially free of noise, Gaussian noise with $\mathrm{rms}=3\times 10^{-9}$~Jy~arcsec$^{-2}$ is added before beam convolution.   In all cubes, the \hi\ disc of the galaxy is set to be inclined by 60~degrees to the y-axis.  MARTINI attempts to identify a preferred disc plane for the galaxy based on the angular momenta of the central 1/3 of particles in the galaxy.  Once an \hi\ data cube is created for a galaxy, it is further smoothed to a spatial resolution of 3.5~arcmin and then has the flux in each channel spatially integrated to generate the \hi\ line profile.  $W_{50}$ is measured as the width of a spectrum at a flux density level equal to half of the peak value.

The stellar masses of the galaxies are taken straight from the \textsc{Caesar} catalogue and are used to calculate the magnitude measure of a galaxy's mean flux surface density within its effective radius, \mueff.  This quantity is used in this work to quantify the degree to which a galaxy's light distribution is centrally concentrated.  As will be explained in Section~\ref{mstarpredicting}, parameterised distributions of \simba\ galaxies in $\log_{10}j_*$~--~$\log_{10}M_*$~--~\mueff\ space are used to predict the derived stellar masses of galaxies taken from the ALFALFA survey and the Spitzer Photometry and Accurate Rotation Curves (SPARC, \citealt{SPARC}) database.  To this end, \mueff\ measurements for the \simba\ galaxies are needed for the SDSS~$I$-band and the Spitzer \irac\ band.  These measurements are henceforth referred to as \mueffi\ and \mueffirac.

To convert a galaxy's mean stellar mass surface density within its effective radius to \mueffi\ and \mueffirac\ in units of apparent magnitudes per arcsec$^2$, an assumed stellar mass-to-light ratio is required for each band.  For the \irac\ band,  0.5~\msun/$L_{\odot}$ is used for all galaxies.  This is the value adopted by \citet{SPARC} to convert total \irac\ luminosities of SPARC galaxies to total stellar masses.  For the SDSS~$I$-band, 1.60~\msun/$L_{\odot}$ is used.   This value comes from considering the absolute magnitudes of all galaxies, taken from PyLoser which computes the dust extinction to each star particle in a galaxy based on its metal content.  The PyLoser magnitudes are converted to luminosities\footnote{Assuming the Vega absolute magnitude of the Sun to be 4.18.} and then compared to the total stellar masses of the galaxies to yield their $I$-band stellar mass-to-light ratios.  For the galaxies used in this study, these ratios are approximately Gaussian-distributed from 0.45 to 2.80~\msun/$L_{\odot}$, with a median value of 1.60~\msun/$L_{\odot}$ and an RMS value of 1.67~\msun/$L_{\odot}$.  Finally, the assumed stellar mass-to-light ratios can be combined with the assumed distance of 4~Mpc for all galaxies to generate measurements of their mean effective surface brightnesses. Most galaxies have \mueffi\ in the range 21 to 23~mag~arcsec$^{-2}$, while 19 to 21~mag~arcsec$^{-2}$ is typical for \mueffirac.  Distributions of these quantities are shown in Figure~\ref{fig:sample}.
   
To generate a realistic uncertainty estimate for each galaxy's stellar specific angular momentum measurement, 8 values of $j_*$ are calculated by considering upper and lower uncertainty limits for each of its constituent factors.  The range of these 8 $j_*$ values is then used as the uncertainty on $j_*$.  To generate an uncertainty range for a galaxy's $W_{50}$ measurement, the half-power width of the \hi\ spectrum is calculated for the peak flux density measured on either side of the galaxy's systemic velocity.  The difference between these two $W_{50}$ measurements is taken as the uncertainty.  An inclination uncertainty of 7.5$^{\circ}$ is used for all galaxies.  The uncertainty on effective radius is set equal to 10~per~cent of the $R_\mathrm{eff}$  measurement taken from the \textsc{Caesar} catalogue. Using this approach, the median relative uncertainty on $j_*$ is 0.19.

\section{Galaxy sample}\label{gal_sample}
\begin{figure*}
	\includegraphics[width=2\columnwidth]{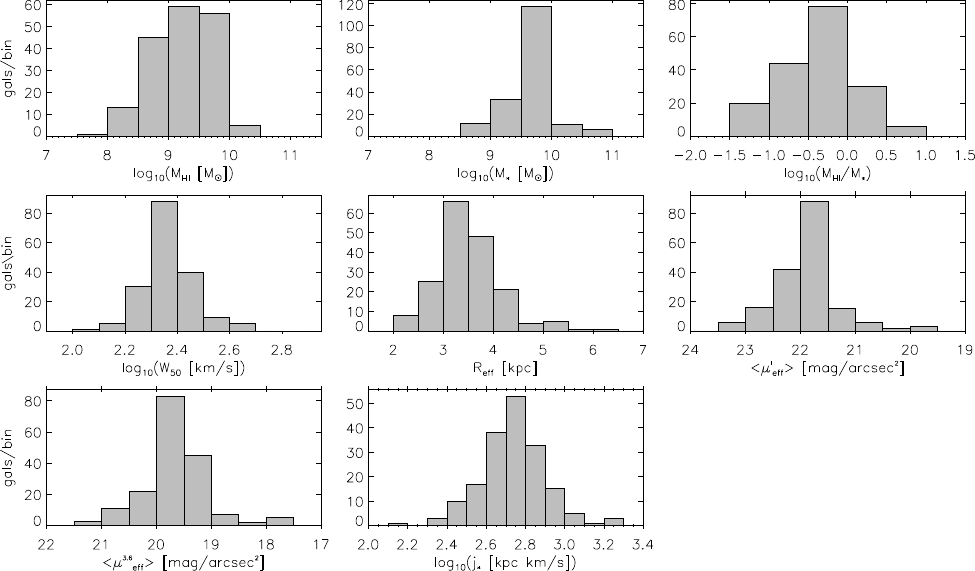}
    \caption{Distributions of various measured and derived properties of the 179 \simba\ galaxies used in this study.  From left to right, top to bottom, the panels show the distributions of HI mass, stellar mass, \hi\ gas fraction, velocity width of the \hi\ profile, stellar disc effective radius, effective surface brightness in the SDSS $I$-band and Spitzer \irac\ band, and stellar specific angular momentum.}
    \label{fig:sample}
\end{figure*}

To generate a sample of \textsc{Simba} galaxies suitable for the current study, various cuts are applied to the full set of 5218 galaxies from the \textsc{Caesar} catalogue. Firstly,  mass cuts are applied: $M_\mathrm{HI}\ge 1.25\times 10^8$~\msun\ and $M_*\ge 7.25\times 10^8$~\msun.  Mass cuts such as these ensure the baryonic mass distributions of the galaxies to be reliably simulated. To ensure galaxies have their dynamics dominated by rotation, a cut on dynamical morphology is made such that the fraction of energy invested in the ordered gas rotation is above 0.75\footnote{The $\mathrm{gas\_kappa\_rot}$ parameter from the \textsc{Caesar} catalogue is used for this.}.  To avoid galaxies with morphologies that are disturbed due to interactions, systems that have galaxy neighbours within a 30~kpc sphere are eliminated.  Additionally, each galaxy has an ellipse fitted to a thin flux annulus in the 12-arcsec version of its \hi\ total intensity map consisting of pixels with mass surface densities in the range 0.9 to 1.1~\msun~pc$^{-2}$.  The mean separation between the pixels and the fitted ellipse is calculated in units of kpc.  Galaxies with a mean separation greater than 5~per~cent the size of the semi-major axis of their ellipse are eliminated.  This further ensures the sample consists of galaxies that have neat, rotating \hi\ discs.  Finally, galaxies with a relative $j_*$ uncertainty greater than 0.25 are removed.  The final sample used in this study consists of 179 \simba\ galaxies.  Figure~\ref{fig:sample} shows the distributions of various measured properties: \hi\ mass, stellar mass, \hi-to-stellar mass, velocity width of \hi\ line profile, stellar disc effective radius, effective surface brightness in the SDSS~$I$-band and Spitzer \irac\ band, and stellar specific angular momentum.

\section{Results}\label{sec:results}
\subsection{\jm\ relation}
\begin{figure}
	\includegraphics[width=\columnwidth]{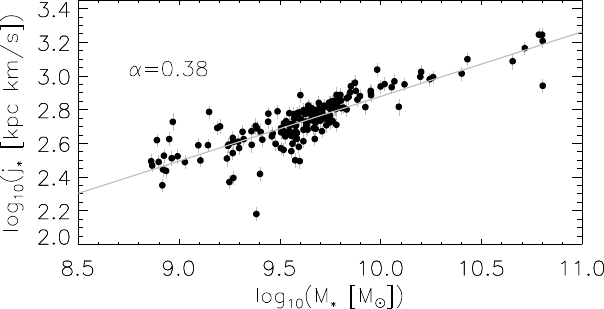}
    \caption{Stellar specific angular momentum as a function of stellar mass for the 179 \simba\ galaxies used in this study.  The solid black line with a slope of 0.38 represents a linear fit to all of the data points.}
    \label{fig:j-M}
\end{figure}

The last panel of Figure~\ref{fig:sample} shows the distribution of $\log_{10}j_*$ measurements for the sample of galaxies used in this study.  Figure~\ref{fig:j-M} shows $\log_{10}j_*$ as a function of $\log_{10}M_*$. Over nearly two decades in stellar mass, the $M_*$ dependence of $j_*$ is well-modelled by a single power law given by
\begin{equation}
\log_{10}\left({j_*\over\mathrm{kpc~km~s^{-1}}}\right) = \alpha\log_{10}\left({M_*\over M_{\odot}}\right)+ \beta,
\end{equation}
with $\alpha = 0.38\pm 0.01$ and $\beta=-0.960\pm 0.162$.  \citet{hardwick_2022a} found $\alpha=0.47\pm 0.02$ for a sample of 564 nearby galaxies from the eXtended GALEX Arecibo SDSS Survey.  More recently, \citet{elson_2024a} measured $\alpha=0.404\pm 0.003$ for a sample of 3607 ALFALFA galaxies spanning the mass range $\sim 10^8$~--~$10^{11}$~\msun.  

\subsection{The $\log_{10}j_*$~--~$\log_{10}M_*$~--~<$\mu_\mathrm{eff}$> plane}
Several authors have shown the \jm\ relation to vary with morphological type and/or with properties that serve as proxies for morphology.  \citet{fall_1983} and \citet{romanowsky_fall_2012} showed spirals and ellipticals to follow parallel \jm\ tracks with slopes close to 0.6, with ellipticals containing roughly 3 to 4 times less angular momentum than spirals of equal mass.  Spatially resolved \hi\ imaging of 16 THINGS galaxies was used by \citet{OG14} to show that bulge-to-disk mass significantly controls the normalisation of the \jm\ relation.  For a large sample of galaxies from the SAMI Galaxy Survey, \citet{cortese_2016} show \jm\ scatter to be strongly correlated with optical morphology (as determined visually and according to  Sérsic index).  More recently, \citet{mancera_pina_2021b} and \citet{hardwick_2022a} showed the \hi\ gas fraction of the interstellar medium to be fundamentally linked to the stellar, baryonic and gas angular momenta of galaxies.

Given that the \textsc{Simba} galaxies used in this study are morphologically diverse\footnote{Consider, for example, the range of \hi\ gas fractions shown in Figure~\ref{fig:sample}.}, the scatter in their \jm\ relation is likely driven significantly by morphology.  Accounting for this should yield tighter, more reliable correlations between $j_*$ and $M_*$.  \citet{elson_2024a} showed the distribution of 3607 ALFALFA galaxies in $\log_{10}j_*$~--~$\log_{10}M_*$~--~\mueffi\ space to be very well modelled by a  2D plane that  yielded $j_*\propto M_*^{0.589\pm 0.02}\left<\mu_\mathrm{eff}\right>^{0.193\pm0.002}$.  The scatter of the galaxies about that best-fitting plane was only 0.089~dex. Thus, the degree to which a galaxy's light is centrally concentrated serves as an effective predictor (in addition to its stellar mass) of its stellar specific angular momentum content.

The primary aim of the current study is to use the \simba\ data to better constrain the planar relation between $\log_{10}j_*$, $\log_{10}M_*$ and \mueff\ found by \citet{elson_2024a}.  A 2D plane of the form 
\begin{equation}\label{eqn:3dplane}
\log_{10}\left({j_*\over \mathrm{kpc~km~s^{-1}}}\right) = \alpha\log_{10}\left({M_*\over M_{\odot}}\right)+\beta\left({<\mu_\mathrm{eff}>\over \mathrm{mag~arcsec^{-2}}}\right)+\gamma
\end{equation}
is fit to the 179 \textsc{Simba} galaxies for which $j_*$, $M_*$ and \mueff\ measurements have been generated in this work.  The IDL function MPFIT2DFUN, which is part of the MPFIT package of curve fitting and function optimisation routines \citep{mpfit}, is used to perform a Levenberg-Marquardt least-squares fit of Equation~\ref{eqn:3dplane} to the data (incorporating the $j_*$ uncertainties discussed at the end of Section~\ref{sec:measurements}).  

As previously mentioned, \mueff\ measurements for the galaxies are produced for the SDSS~$I$-band and Spitzer \irac\ band by assuming a constant mass-to-light ratio in each band.  Therefore, the distributions of the \simba\ galaxies in $\log_{10}j_*$~--~$\log_{10}M_*$~--~\mueffi\ space and $\log_{10}j_*$~--~$\log_{10}M_*$~--~\mueffirac\ space are exactly parallel to one another, offset only along the \mueff\ axis. For both distributions, $\alpha = 0.694\pm0.046$ and $\beta=0.190\pm 0.026$ are the best-fitting plane parameters.  The distribution of measured perpendicular distances between the galaxies and either of the two best-fitting planes is very well approximated by a Gaussian, with a median separation of 0.007 dex and an RMS separation of only 0.057 dex.  The best-fitting $\gamma$ parameters for the $I$-band and \irac\ band planes are $-8.111\pm 1.009$ and $-7.691\pm 0.953$, respectively.  

While the best-fitting $\beta$ and $\gamma$ parameters found for the \textsc{Simba} galaxies (for the \mueffi\ case) are well-consistent within 1$\sigma$ of the best-fitting values from \citet{elson_2024a}, the $\alpha$ value found in the current study is consistent within $\sim 2.5\sigma$.  Compared to $\alpha=0.58\pm 0.002$ from \citet{elson_2024a}, the best-fitting $\alpha$ parameter for the \textsc{Simba} galaxies is $\sim 20$~per~cent higher.  This could be due to uncertainties of up to a few tenths of a dex on the stellar masses from \citet{durbala_2020} that were used to measure $\alpha$ in the ALFALFA study.  It could indicate that the approximation used for a galaxy's $j_*$ content (in both studies) might be less accurate than expected under certain conditions.  The 3607 ALFALFA galaxies from the \citet{elson_2024a} study likely have the properties of their \hi\ discs being affected by environmental factors in ways that the \simba\ galaxies from the present study do not.  However, this study's best-fitting $\alpha$ parameter is very close to the value of $\alpha=0.67\pm 0.03$ from \citet{mancera_pina_2021b} who fitted planes to the distributions of galaxies in $\log_{10}j_*$~--~$\log_{10}M_*$~--~$\log_{10}f_\mathrm{gas}$, where $f_\mathrm{gas}$ is that ratio of total gas mass to total baryonic mass.  It is also very close to the theoretical expectation of $\alpha=2/3$ for galaxies in a $\Lambda$CDM cosmology\footnote{See \citealt{posti_2018b} for a derivation of the relevant equation showing $j_*\propto M_*^{2/3}$.}. The very small amount of scatter (0.057~dex) of the \simba\ galaxies about the best-fitting plane suggests it represents a fundamental relationship between $j_*$, $M_*$ and \mueff, at least for late-type galaxies dominated by rotation.  To test this idea, the following section is dedicated to using the plane to predict the stellar masses of real galaxies. 

\section{Predicting stellar masses}\label{mstarpredicting}
Given the very tight planar distributions of the \simba\ galaxies in $\log_{10}j_*$~--~$\log_{10}M_*$~--~\mueff\ space, this study's parameterisations of the best-fitting planes for the \mueffi\ and \mueffirac\ cases  can be used to accurately predict the stellar masses of real galaxies for which there exist $j_*$ and \mueff\ measurements (determined independently of $M_*$). 

\subsection{ALFALFA}
The stellar masses of the 3607 ALFALFA galaxies from \citet{elson_2024a} are predicted using their $j_*$ and \mueffi\ measurements from that study, and compared to the stellar masses from \citet{durbala_2020}.

\begin{figure*}
	\includegraphics[width=2\columnwidth]{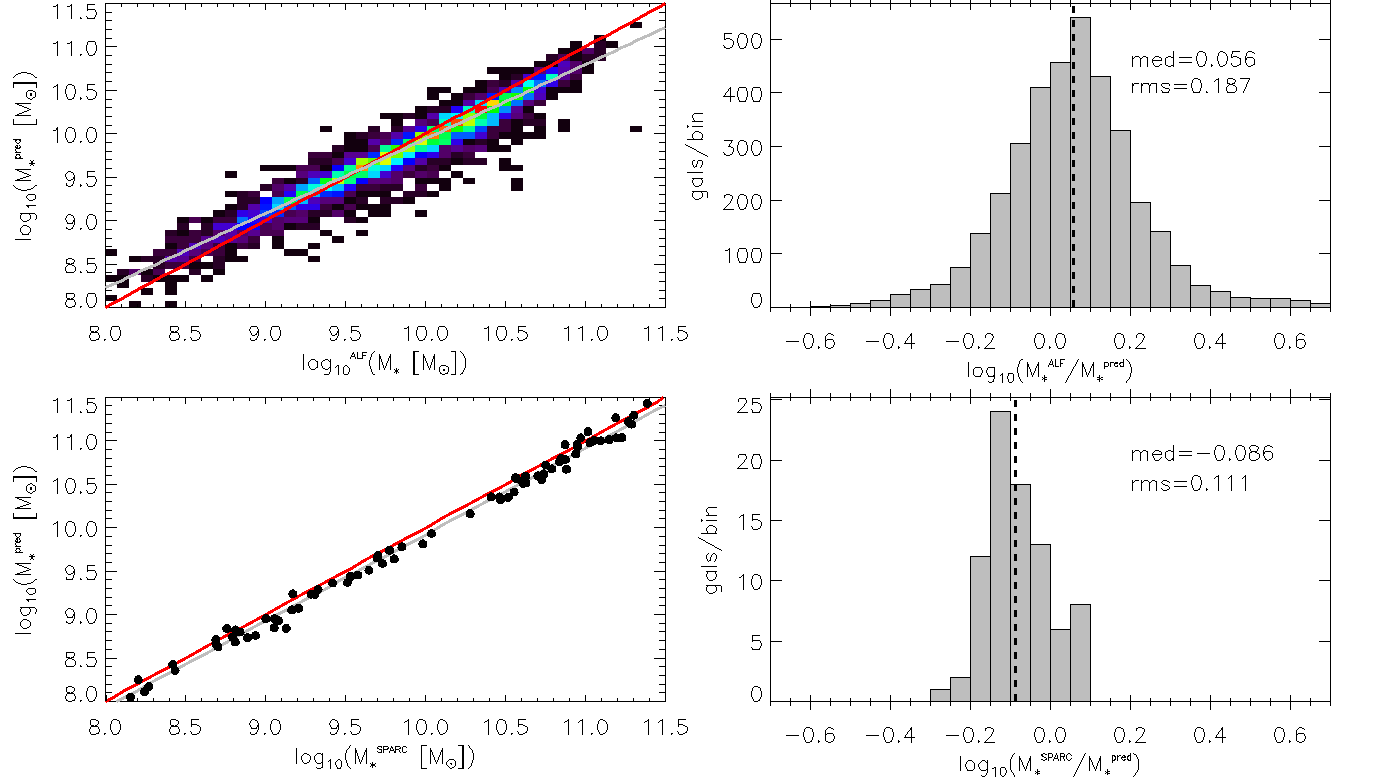}
    \caption{Comparisons between predicted and derived stellar masses for ALFALFA galaxies (top row) and SPARC galaxies (bottom row).  Left panels: predicted stellar masses as a function of derived stellar mass.  Grey lines represent a first-order polynomial fit to the data, red lines represent equality.  The 2D histogram for the ALFALFA galaxies is based on square-shaped bins of side length 0.075 dex.  Right panels: distribution of differences between predicted and derived stellar masses.  Shown in each panel are the median and RMS of the distribution.  These results demonstrate the best-fitting planes from this study to serve as effective tools for accurately predicting the stellar masses of real galaxies.}
    \label{fig:Mstar_predict_ALFALFA}
\end{figure*}

Figure~\ref{fig:Mstar_predict_ALFALFA} shows in its top-left panel the 2D distribution of the predicted $M_*$ values\footnote{Based on $\alpha =  0.694$, $\beta=0.190$, $\gamma=--8.111$.} as a function of those from the \citet{durbala_2020} catalogue. The solid grey line is a fit to the median predicted $M_*$ values in bins of width 0.1 dex along the x-axis, while the red line is that of equality.  The predictive ability of this study's fitted plane is seen to vary with stellar mass.  While galaxies with derived stellar masses close to $\sim10^{9.5}$~\msun\ are very well predicted, masses are under/over predicted by up to a few tenths of a dex at lower/higher masses.  For all 3607 galaxies, the top-right panel in Figure~\ref{fig:Mstar_predict_ALFALFA} shows the distribution of the differences between predicted and derived stellar masses.  Over $\sim 3$ decades of stellar mass, the median discrepancy is 0.056~dex while the RMS discrepancy is 0.187 dex.  

\begin{figure*}
	\includegraphics[width=2\columnwidth]{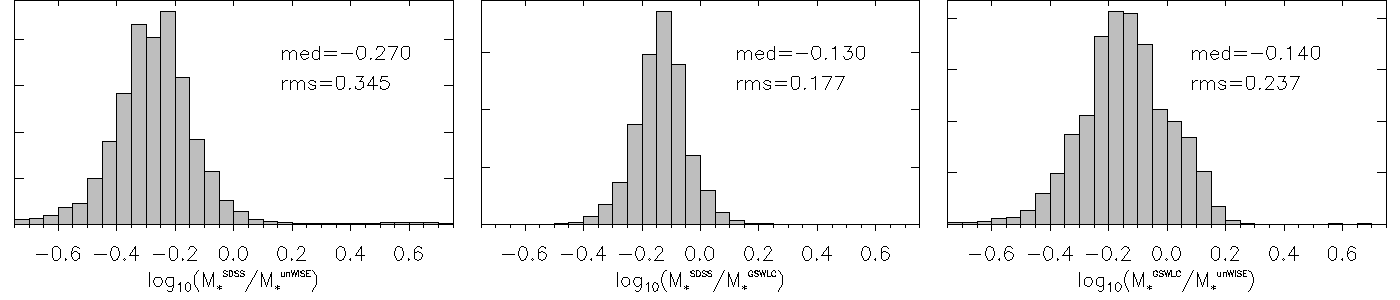}
    \caption{For the 3607 ALFALFA galaxies used in \citet{elson_2024a}, the panels show differences in logarithmic stellar masses based on SDSS optical photometry and infrared unWISE photometry and/or ultraviolet imaging from GALEX (left panel), SDSS optical photometry and measurements from GALEX-SDSS-WISE Legacy Catalog 2 (middle panel), measurements from GALEX-SDSS-WISE Legacy Catalog 2 and infrared unWISE photometry and/or ultraviolet imaging from GALEX (right panel).  Differences in derived stellar masses from these three sources are typically larger than the error with which the current study's parameterisation of the distribution of \simba\ galaxies in $\log_{10}j_*$~--~$\log_{10}M_*$~--~\mueff\ space can predict them.}
    \label{fig:Mstar_compare}
\end{figure*}

The above-mentioned discrepancies are small compared to uncertainties typically associated with derived stellar masses. Figure~\ref{fig:Mstar_compare} considers the ALFALFA galaxies from \citet{elson_2024a} that have more than one derived stellar mass available and shows the discrepancies between them.  These masses are all taken from the \citet{durbala_2020} catalogue and are derived either from SDSS optical photometry, infrared unWISE photometry and/or ultraviolet imaging from GALEX \citep{mcGaugh_2015}, or multi-wavelength spectral energy distribution fitting from the GALEX-SDSS-WISE Legacy Catalog 2 \citep{salim_2016, salim_2018}.  Very clear is the fact that all distributions are well-approximated by a Gaussian with a mean significantly offset from zero and with RMS ranging from $\sim 0.18$ to $\sim 0.35$~dex.  Thus, this study's parameterisation of the distribution of \simba\ galaxies in $\log_{10}j_*$~--~$\log_{10}M_*$~--~\mueffi\ space is generally able to predict the stellar masses of real galaxies with more reliability and less uncertainty than the differences between the various derived stellar masses that are based on sophisticated modelling processes.  Any SDSS galaxy with available $W_{50}$ and inclination measurements can have its stellar masses accurately predicted. 

\subsection{SPARC galaxies}
As a second test of the predictive ability of this study's results, galaxies from the database of the Spitzer Photometry and Accurate Rotation Curves (SPARC, \citealt{SPARC}) are considered. The data gathered in Table1.mrt on the SPARC website\footnote{http://astroweb.cwru.edu/SPARC/} is used to generate $j_*$ measurements for the SPARC galaxies:
\begin{equation}
j_*^\mathrm{SPARC} = 2V_\mathrm{flat}R_\mathrm{eff}/1.68,
\end{equation}
where $V_\mathrm{flat}$ is the asymptotically flat rotation velocity in units of \kms\ and $R_\mathrm{eff}$ is the effective radius in units of kpc as measured in the \irac\ band of the Spitzer telescope. A \mueffirac\ measurement for each SPARC galaxy comes from using half of its total luminosity at \irac\ and dividing it by the area within its effective radius, and then taking into account its distance.  Only those SPARC galaxies with a quality flag of 1 are considered.  Additionally, the relative uncertainty in $V_\mathrm{flat}$ must be less than 0.1.  These cuts yield a sample of 84 SPARC galaxies.

\citet{SPARC} convert total \irac\ luminosity in $L_{\odot}$ units to total stellar mass in \msun\ units by assuming a constant stellar mass-to-light ratio of 0.5~$M_{\odot}/L_{\odot}$.  These stellar masses are compared to those predicted using this study's 2D plane together with its $j_*$ and \mueffirac\ measurements for the SPARC galaxies. Figure~\ref{fig:Mstar_predict_ALFALFA} shows in its bottom-left panel the predicted $M_*$ values\footnote{Based on $\alpha =  0.694$, $\beta=0.190$, $\gamma=-7.691$.} as a function of those based on the information from the SPARC website.  The grey line represents a linear fit to the data while the red line is that of equality.  Very clear is the fact that the current study's best-fitting $\log_{10}j_*$~--~$\log_{10}M_*$~--~\mueffirac\ plane has a very high level of predictive accuracy that is constant with stellar mass. The bottom-right panel of Figure~\ref{fig:Mstar_predict_ALFALFA} shows the difference to be approximately Gaussian distributed with a median value of -0.086~dex and an RMS of only 0.111~dex.  These results again demonstrate the best-fitting planes from this study to serve as effective tools for accurately predicting the stellar masses of real galaxies.

\section{Summary}\label{sec:summary}
This work aims to use galaxies from the \simba\ simulation to study their distribution in $\log_{10}j_*$~--~$\log_{10}M_*$~--~\mueff\ space to better constrain the dependence of stellar specific angular momentum ($j_*$) on stellar mass ($M_*$) and the central concentration of a galaxy's light - as quantified by the magnitude measure of the average flux within its effective radius (\mueff). 

Spanning a stellar mass range $8.5\lesssim\log_{10}(M_*/M_\odot)\lesssim 11$, 179 \simba\ galaxies are selected to have their $j_*$, $M_*$ and \mueff\ quantities measured.  A galaxy's gas particles are used to produce an \hi\ data cube from which a measure of the velocity width of its \hi\ line profile is extracted.  The star particles of a galaxy are used to directly acquire measurements of disc scale length and total stellar mass.  These measurements are all combined to approximate $j_*$.

The distribution of \simba\ galaxies in $\log_{10}j_*$~--~$\log_{10}M_*$~--~\mueff\ space is found to be very well-approximated by a 2D plane that yields $j_*\propto M_*^{0.694} \mu_\mathrm{eff}^{0.190}$.  This measured mass dependence of $j_*$ on $M_*$ is very close to what is predicted by $\Lambda$CDM theory.  The RMS value of the perpendicular distances of the galaxies about the best-fitting plane is only 0.057~dex, thereby making the planar distribution of \simba\ galaxies much tighter than the distribution of ALFALFA galaxies from \citet{elson_2024a}.  

The utility of the parameterised planar distribution of \simba\ galaxies in $\log_{10}j_*$~--~$\log_{10}M_*$~--~\mueff\ space as a tool for accurately predicting the stellar masses of real galaxies with SDSS~$I$-band or Spitzer~\irac\ photometry is demonstrated.  Over a large mass range, derived stellar masses of galaxies from the ALFALFA survey and the database of the Spitzer Photometry and Accurate Rotation Curves (SPARC) are accurately predicted to within $\sim 0.1$ to 0.2~dex.  Considering the large number of ongoing \hi\ and optical/infrared galaxy surveys, the planar relations presented in this study make it possible to accurately estimate the stellar masses of their detected galaxies without the need for detailed modelling and acquisition of additional multi-wavelength imaging. 

This study serves as a theoretical demonstration of the intrinsic relationship between stellar specific angular momentum, mass and morphology for late-type galaxies, and presents yet another galaxy scaling relation from the \simba\ simulations that is highly consistent with empirical results. 

\section*{Acknowledgements}
Sincere thanks are extended to the anonymous referee for providing truly insightful comments that improved the quality of this work. 

\section*{Data Availability}
Upon reasonable request, the author is willing to make available the measurements generated in this study.



\bibliographystyle{mnras}


%


\bsp	
\label{lastpage}
\end{document}